\documentclass{ws-procs9x6square}

\def\al{\alpha}
\def\be{\beta}
\def\ga{\gamma}
\def\de{\delta}

\def\et{\eta}

\def\la{\lambda}

\def\fr#1#2{{{#1} \over {#2}}}

\def\frac#1#2{{\textstyle{{#1}\over {#2}}}}

\def\lsim{\mathrel{\rlap{\lower4pt\hbox{\hskip1pt$\sim$}}
    \raise1pt\hbox{$<$}}}
\def\gsim{\mathrel{\rlap{\lower4pt\hbox{\hskip1pt$\sim$}}
    \raise1pt\hbox{$>$}}}
\def\sqr#1#2{{\vcenter{\vbox{\hrule height.#2pt
         \hbox{\vrule width.#2pt height#1pt \kern#1pt
         \vrule width.#2pt}
         \hrule height.#2pt}}}}

\def\lrpartial{\raise 1pt\hbox{$\stackrel\leftrightarrow\partial$}}

\def\etal{{\it et al.}}

\newcommand{\beq}{\begin{equation}}
\newcommand{\eeq}{\end{equation}}
\newcommand{\bea}{\begin{eqnarray}}
\newcommand{\eea}{\end{eqnarray}}
\newcommand{\rf}[1]{(\ref{#1})}

\begin{document}


\title{Lorentz-violating dispersion relations and threshold analyses}

\author{Ralf Lehnert}
\address{CENTRA, Departamento de F\'{\i}sica\\
Universidade do Algarve\\
8000-117 Faro\\
Portugal\\
E-mail: rlehnert@ualg.pt}


\maketitle

\abstracts{
Remnant low-energy effects of Planck-scale Lorentz breaking
in candidate fundamental theories
typically include modified one-particle dispersion relations.
Theoretical constraints
on such modifications
are discussed
leading to the exclusion
of a variety of previously considered
Lorentz-violating parameters.
In particular,
the fundamental principle of coordinate independence,
the role of an effective dynamical framework,
and the conditions of positivity and causality
are investigated.
}


\section{Introduction}

An important open issue
in current fundamental-physics research
is a quantum theory underlying the Standard Model
and General Relativity.
The characteristic scale of such a theory
is likely to be associated
with the Planck mass $M_{\rm Pl}\simeq 10^{19}$ GeV.
Presently attainable energies
are minuscule compared to this scale,
so that experimental signals
are expected to be heavily suppressed.
For observational progress in this subject,
it is therefore necessary
to identify generic effects of potential fundamental theories
that are accessible to high-precision tests
with present-day technology.

Relativity violations
associated with the breaking Lorentz symmetry
provide a promising candidate signal
for Planck-scale physics \cite{cpt01}.
At low energies,
the effects of Lorentz violation
are described by an effective field theory
called the Standard-Model Extension (SME) \cite{ck97,kl01}.
The classical action of the SME
contains,
for example,
all leading-order contributions to the Lagrangian
formed by contracting Standard-Model and gravitational fields
with Lorentz-breaking coefficients
such that coordinate independence is maintained.
A theoretically attractive mechanism for Lorentz violation
is spontaneous symmetry breaking
in string field theory \cite{kps}.
More recently,
also other candidate sources
have been considered
including
theories of quantum gravity \cite{foam},
noncommutative field theories \cite{ncft},
varying couplings \cite{vc},
random dynamics \cite{rd},
multiverses \cite{mv},
and brane-world scenarios \cite{bws}.
The flat-spacetime limit of the SME
has provided the basis for
numerous analyses of Lorentz breaking
involving mesons \cite{hadronexpt,hadronth},
baryons \cite{ccexpt,kl99},
electrons \cite{eexpt,bkr},
photons \cite{photonexpt,photonth},
muons \cite{muons},
and neutrinos \cite{ck97,cg99,neutrinos}.

One-particle dispersion relations
extracted from the SME
generally exhibit Lorentz-violating modifications \cite{photonexpt,ck97,kl01}.
In principle,
this offers the possibility of Lorentz tests
with purely kinematical methods.
For instance,
primary ultrahigh-energy cosmic rays (UHECR)
with momenta
eight orders of magnitude
below the Planck scale
have been observed.
At such energies,
Lorentz-breaking effects
might be more pronounced
relative to those in low-energy tests
leading to potentially observable
shifts in particle-reaction thresholds.
This idea
has recently received a lot of attention in the literature
\cite{foam,cg99,thres,rl03}.
However,
in many of these investigations
the dispersion-relation modifications
are constructed with a certain degree of arbitrariness
and without reference to the underlying dynamics
and other physical principles.

In this talk,
we investigate
how some of this arbitrariness
can be avoided.
Our analysis
is primarily based
on the fundamental principle of coordinate independence
and on the requirement of compatibility
with an effective dynamical framework.
We argue
that these two conditions
form cornerstones of physics
regardless of the details of the Planck-scale theory.
General dynamical considerations
also increase the scope of threshold investigations
and may even be necessary in certain situations.
In addition,
we briefly discuss positivity and causality,
properties that further contribute
to the viability of kinematical studies.
Throughout we assume translational invariance
and the associated energy-momentum conservation.

Section \ref{coordind}
comments on the necessity of coordinate independence
and its consequences for dispersion relations.
In Sec.\ \ref{smesec},
we discusses
dispersion relations
from the viewpoint of compatibility
with the SME.
Section \ref{poscaus}
addresses issues regarding positivity and causality.
Further useful results
are contained in Sec.\ \ref{further}.
A brief summary can be found in Sec.\ \ref{sum}.

\section{Coordinate independence}
\label{coordind}

On the one hand,
coordinate independence
is a fundamental physics principle,
its role in the context of Lorentz breaking
is well established \cite{ck97,rl03},
and it permits a rough classification
of different types of Lorentz violation.
On the other hand,
there still exists a certain amount of confusion
about this principle in the published literature.
For instance,
dispersion-relation corrections
considered by some authors
can only be reconciled with coordinate invariance
by introducing unsatisfactory features.
Occasionally Lorentz violation is even identified
with the loss of coordinate independence.
We therefore begin with a few remarks
about this requirement.

Coordinate systems,
which label spacetime points in a largely arbitrary way,
are descriptive tools
rather than objects with physical reality.
Physics must therefore remain independent of the choice of coordinates.
This fundamental requirement
permits different observers,
each describing the {\it same} physical system
within a {\it different} reference frame,
to relate their observations.
This principle is therefore also called
observer invariance.
Mathematically,
coordinate independence can be implemented
by working on a spacetime manifold
and representing physical quantities
by geometric objects like tensors or spinors.
Note,
however,
that this principle
does not fix the type of the underlying manifold.
A Lorentzian and a Galilean manifold,
for example,
would be equally consistent with coordinate independence.
The manifold type
can only be determined by observation.
The point is
that coordinate independence
is much more general
than Lorentz symmetry.
It is only on Lorentzian manifolds
where Lorentz transformations acquire the significant role
of implementing changes between local Minkowski frames.

The above discussion reveals one possibility
to speculate how Lorentz symmetry might be lost:
local inertial frames have a structure
different from the usual Minkowskian one,
so that Lorentz transformations
no longer generate changes between inertial coordinates,
i.e.,
observer Lorentz covariance is {\it replaced}
by observer covariance
under some other symmetry transformation.
Note
that coordinate independence is maintained.
This point of view is taken
in the so called ``doubly special relativities''
\cite{dsr}.
The associated modified dispersion relations
still exhibit the conventional energy degeneracy
for a given 3-momentum,
which is intuitively reasonable
because the number of spacetime symmetries remains unchanged
relative to the conventional case.
We mention
that the interpretation and viability of this approach
is currently still controversial \cite{triv}.
We therefore leave
such Lorentz-symmetry deformations unaddressed in the present work.

A less speculative approach to Lorentz-symmetry breakdown
maintains the conventional Lorentzian manifold structure
and considers the vacuum to be nontrivial instead.
Such vacua are associated
with nondynamical tensorial backgrounds,
which can lead,
e.g.,
to direction-dependent propagation properties.
This situation has some parallels
with the behavior of particles
inside certain crystals.
Although coordinate independence is maintained
(e.g., invariance under rotations of the {\it coordinate system}),
rotations of the {\it propagation direction}
are generally no longer a symmetry in such situations.
One then says
that {\it particle Lorentz symmetry} is broken \cite{ck97}.
Note, however,
that the presence of the conventional manifold structure implies
that locally one can still work
with the metric $\et^{\mu\nu}={\rm diag}(1,-1,-1,-1)$,
particle 4-momenta
still transform in the usual way under coordinate changes,
and the conventional tensors and spinors
still represent physical quantities.
In what follows,
we consider this latter type of Lorentz violation.

The usual form of Lorentz-violating dispersion relations
considered in the literature is\footnote{Dispersion relations
in the minimal SME are typically fourth-order polynomials
in $\la_0$,
so that \rf{lvdr} together with ansatz \rf{ansatz} is inconvenient.
However,
SME dispersion relations can be generated
if $\de f$ is allowed to contain unsuppressed terms.
}
\beq
{\la_0}^2-{\vec{\la}}^{\:2}=m^2+\de f(\la_0,\vec{\la})\; .
\label{lvdr}
\eeq
Here, $m$ is the mass
and $\la^{\mu}=(\la^0,\vec{\la})$
is the plane-wave 4-vector
(before the QFT reinterpretation of the negative energies).
The function $\de f(\la_0,\vec{\la})$
parametrizes the Lorentz breaking.
Arbitrary choices of $\de f(\la_0,\vec{\la})$
include situations with nonconserved
and possibly complex-valued momenta unsuitable
for kinematical analyses.\footnote{For instance,
a discrete background,
such as in condensed-matter systems,
lacks translation invariance
resulting in the violation of momentum conservation.
Another example is given by finite-temperature dispersion relations,
which typically contain imaginary terms.}
We exclude such situation here
and proceed under the assumption
that the dynamics of the free particle is described
by a linear partial differential equation
with constant coefficients, as usual.
In the absence of nonlocalities,
this yields a polynomial ansatz:
\beq
\de f(\la_0,\vec{\la})=
\sum_{n\ge 1}\hspace{4mm}
\overbrace{\hspace{-3.5mm}T_{(n)}^{\;\al\be\;\cdots\;\;}}^{n\; \rm indices}
\hspace{-1mm}\underbrace{\hspace{.5mm}\la_{\al}\la_{\be}\;\cdots\;}
_{n\; \rm factors} \;\; .
\label{ansatz}
\eeq
Here, $T_{(n)}^{\;\al\be\;\cdots\;}$ is a constant tensor of rank $n$
parametrizing particle Lorentz violation.
All the tensor indices $\al,\be,\,\ldots$
are distinct
and each one is contracted with a momentum factor,
so that all terms in the sum are observer Lorentz invariant.

We mention two immediate consequences
of the general ansatz \rf{ansatz}.
First,
Eq.\ \rf{lvdr} becomes a polynomial in $\la_0$,
so that one expects multiple roots
for a given $\vec{\la}$.
Thus, the conventional energy degeneracy
between particle, antiparticle, and possible spin-type states
is typically lifted.
This is intuitively reasonable
because degeneracies normally arise through symmetries,
and here the number of symmetries
is reduced.
Moreover,
rotational invariance is in general lost in any frame.
As opposed to all previous threshold analyses,
generality therefore requires the consideration
degeneracy-lifting anisotropic dispersion relations.
Second,
inspection of ansatz \rf{ansatz} shows
that under the usual assumption of rotational symmetry
the correction $\de f(\la_0,\vec{\la})$
cannot contain odd powers of $|\vec{\la}|$.\footnote{Odd powers
could enter in the degeneracy-lifting form $\pm|\vec{\la}|^{2n+1}$
with $n\in {\mathbb N}$
or through
(theoretically unmotivated)
nonlocal equations of motion involving $\sqrt{-\Delta}$,
where $\Delta$ is the Laplacian.}

\section{Dynamical features}
\label{smesec}

Although kinematics imposes tight constraints on particle reactions,
it provides only an incomplete description of the process:
an expected high-energy reaction can be suppressed
not only by modified dispersion relations
but also by new additional symmetries, for example.
Similarly,
the presence of a high-energy reaction
kinematically forbidden in conventional physics
could perhaps be explained by additional channels due to the loss of
low-energy symmetries or novel undetected particles.
Moreover,
dynamics
is involved both
in acceleration mechanisms for UHECRs
and in the atmospheric shower development.
Thus,
the study of threshold bounds on Lorentz violation
typically requires assumptions outside kinematics
such as dynamical quantum-field aspects.

Kinematics investigations
are limited to only a few
potential Lorentz-violating signatures
from candidate fundamental physics.
Thus,
dynamical features also
increase the scope of Lorentz tests.
From the above perspectives,
it is desirable
to explicitly implement dynamics of sufficient generality
into the search for Lorentz breaking.

The SME is the general effective-field-theory framework
for the dynamical description of Lorentz violation.
It is useful to review the idea behind its construction \cite{ck97}
to fully appreciate the generality of the SME.
Lorentz-violating terms $\de {\mathcal L}$
are added to the usual Standard-Model Lagrangian ${\mathcal L}_{\rm SM}$:
\beq
{\mathcal L}_{\rm SME}={\mathcal L}_{\rm SM}+\de {\mathcal L}\; ,
\label{sme}
\eeq
where, ${\mathcal L}_{\rm SME}$ denotes the SME Lagrangian.
The modification $\de {\mathcal L}$
is formed by contracting Standard-Model fields
with Lorentz-violating tensorial parameters
yielding observer Lorentz scalars.
Thus,
the complete set of possible contributions to $\de {\mathcal L}$
gives the most general effective dynamical framework
for Lorentz breaking
at the level of Lorentz-coordinate-independent effective QFT.
We mention
that potential Planck-scale features,
such as a certain  discreteness of spacetime
or a possible non-pointlike nature of elementary particles,
are unlikely to invalidate
the above effective-field-theory approach
at present energies.
Moreover,
Lorentz-symmetric aspects
of candidate fundamental theories,
such as new symmetries,
novel particles,
or large extra dimensions,
are also unlikely to require a low-energy description
beyond effective field theory
and can therefore be incorporated into the SME,
if necessary.

The SME
permits the identification
and direct comparison
of virtually all Lorentz and CPT tests
that are presently feasible.
In addition,
classical kinematics test models of relativity
(such as Robertson's framework,
its Mansouri-Sexl extension,
or the $c^2$ model)
are contained in the SME
as limiting cases.
Concerning threshold analyses,
the quadratic, translationally invariant sector
of the SME
determines possible one-particle dispersion relations
constraining the ansatz \rf{ansatz}.
As a further advantage,
the SME permits the calculation of reaction rates,
which are a determining factor for observational relevance.
An explicit example is provided
by so called vacuum \v{C}erenkov radiation \cite{cg99}.
The above discussion strongly suggests
that particle-reaction investigations
are best performed within the framework of the SME.

\section{Positivity and causality}
\label{poscaus}

In Special Relativity,
the presence of an upper speed limit
for material bodies
left invariant by the Lorentz transformations
is associated with a notion of causality.
This leads to the common misconception
that Lorentz violation
necessarily results in superluminal propagation,
and thus causality violations.
However,
conventional situations,
in which Lorentz symmetry is broken
but causality is maintained,
can readily be identified.
The anisotropic propagation of electromagnetic waves
inside certain crystals,
for example,
is causal despite Lorentz violation.
Moreover,
in such a situation the total conserved energy
is clearly positive definite for all observers.
It follows
that the requirements of positivity and causality
are {\it a priori} independent and distinct
from the principle of Lorentz symmetry.
Note also
that positivity and causality
lead, for example, to the spin-statistics theorem,
which is a cornerstone of relativistic QFT.

Since polynomial Lorentz-violating dispersion relations
can violate positivity and causality \cite{kl01},
it is natural to ask
whether such violations
become acceptable in the presence of Lorentz breaking.
Concerning positivity,
we are unaware
of any internally consistent interacting quantum field theories
involving negative-energy particles as asymptotic states.
On the contrary,
the usual assumptions in perturbation theory,
for example,
seem to exclude negative energies.
Similar arguments apply to superluminal propagation:
it is unlikely
that such a causality breakdown
can be accommodated within the framework
of relativistic quantum field theory.
Generally,
a hermitian Hamiltonian for massive fermions
fails to exist in the majority
of frames \cite{kl01}.
In addition,
the usual covariant perturbative expansion
relies on time ordering,
an operation no longer coordinate invariant
when microcausality is violated \cite{green02}.
We conclude
that positivity and causality
remain desirable features in threshold analyses
despite Lorentz breaking.

Reaction-threshold kinematics can be affected
if positivity and causality are imposed.
Let $M$ and $m$ be the respective scales
of the underlying theory and present-day low-energy physics.
Then,
the scale $p_{p{\textrm -}c}$ for the occurrence of positivity
or causality problems
can be as low as \cite{kl01}
\beq
p_{p{\textrm -}c}\sim{\mathcal O}(\sqrt{mM}\:)\; .
\label{scale}
\eeq
For example,
if $M$ is taken to be the Planck scale
and $m$ is the proton mass,
then $p_{p{\textrm -}c}\sim3\times 10^{18}$ eV.
UHECRs with a spectrum extending beyond $10^{20}$ eV
have been observed.
These events
are often employed to bound Lorentz breaking
or to suggest evidence for Lorentz violation.
It follows
that imposing positivity and causality
could require modifications in threshold analyses.

\section{Further results}
\label{further}

Consider photon decay $\ga\rightarrow e^+ + e^-$
into an electron-positron pair,
where both the photon ($m=0$) and the fermion
obey dispersion relations with the correction
$\de f(\la_0,\vec{\la})=\pm |\vec{\la}|^3/M$.
Here, $M$ is the fundamental scale.
Note
that by allowing two simultaneous signs for the correction term,
we enforce coordinate independence.
This correction gives
\beq
\la^0_{\pm (\al)}(\vec{\la})=\pm\sqrt{(-1)^{\al}
\fr{|\vec{\la}|^{\,3}}{M}+\vec{\la}^{\:2}+m^2}\; ,
\label{eigenen3}
\eeq
where
the subscript $\pm$ corresponds to the sign of the square root,
and thus,
after reinterpretation,
to particle and antiparticle.
The index $\al=1,2$ labels the two possible particle (antiparticle) energies,
which perhaps correspond to different spin-type states.
Depending on the $\al$ value for each particle in the reaction,
there are six kinematically distinct decays
that have to be considered.
Note, however,
that angular-momentum conservation
associated with the rotational invariance of the model
may preclude some of the six reactions.
In general,
we conclude
that the effects of assumed symmetries,
such as rotational invariance,
must be incorporated into threshold analyses.
This typically requires the use of dynamics
as argued before.

Another coordinate-independent dispersion-relation correction
is given by $\de f(\la_0,\vec{\la})=\la_0\vec{\la}^{\:2}/M$,
so that
\beq
\la^0_{\pm}(\vec{\la})=\fr{\vec{\la}^{\:2}}{2M}
\pm\sqrt{\fr{\vec{\la}^{\:4}}{4M^2}+\vec{\la}^{\:2}+m^2}\; .
\label{eigenen2}
\eeq
Note that the particle-antiparticle degeneracy is lifted.
Consider again photon decay $\ga\rightarrow e^+ + e^-$,
now with dispersion relation \rf{eigenen2}
for both the photon ($m=0$) and the fermion.
Two kinematically distinct processes must be investigated
because Eq.\ \rf{eigenen2}
implies two possible incoming photon states $\ga_+$ and $\ga_-$.
One can show \cite{rl03}
that the reaction $\ga_{-}\rightarrow e^++e^-$
is kinematically forbidden,
while the decay $\ga_{+}\rightarrow e^++e^-$ is allowed
above a certain threshold.
This analysis has been performed previously in the literature
employing the approximation $\la_0\simeq|\vec{\la}|$
in the correction term $\la_0\vec{\la}^{\:2}/M$.
However,
this approximation introduces an additional degeneracy
relative to Eq.\ \rf{eigenen2}
leading to the false conclusion
that the correction $\la_0\vec{\la}^{\:2}/M$
precludes photon decay.
Thus,
many approximations,
such as those leading to additional degeneracies,
are typically invalid in threshold analyses.

\section{Summary}
\label{sum}

This talk has discussed
some issues
that arise in the context of
Lorentz tests with modified dispersion relations.
More specifically,
we have investigated
the role of a dynamical framework
and the conditions
of coordinate independence,
positivity,
and causality
in the subject.
We have found
that these fundamental requirements
impose tight constraints
on possible dispersion-relation corrections.
Correct threshold investigations within the full SME
are automatically compatible with these requirements.

\section*{Acknowledgments}

This work was supported in part
by the Centro Multidisciplinar de Astrof\'{\i}sica (CENTRA)
and by the Funda\c{c}\~ao para a Ci\^encia e a Tecnologia (Portugal)
under grant POCTI/FNU/49529/2002.

\appendix

\end{document}